\documentclass[12pt,eqno,epsf]{article}
\usepackage{amsmath,amssymb,graphicx,slashbox}
\numberwithin{equation}{section}

\def\ignore#1{{}}

\newcounter{sxn}

\newcounter{axn}

\date{}

\newdimen\mybaselineskip
\renewcommand{\thefootnote}{\arabic{footnote}}

\newcommand{\beeq}{\begin{equation}}
\newcommand{\eneq}{\end{equation}}
\newcommand{\beqn}{\begin{eqnarray}}
\newcommand{\eeqn}{\end{eqnarray}}

\newcommand{\alp}{\alpha}

\newcommand{\dlt}{\delta}

\newcommand{\tht}{\theta}

\newcommand{\vth}{\vartheta}
\newcommand{\kp}{\kappa}
\newcommand{\lmd}{\lambda}

\newcommand{\sgm}{\sigma}
\newcommand{\Sgm}{\Sigma}

\newcommand{\vph}{\varphi}
\newcommand{\omg}{\omega}
\newcommand{\Omg}{\Omega}

\newcommand{\be}{\begin{equation}}
\newcommand{\ee}{\end{equation}}
\newcommand{\bea}{\begin{eqnarray}}
\newcommand{\eea}{\end{eqnarray}}
\newcommand{\eql}{\!\!\!&=\!\!\!&}

\newcommand{\sma}{\!\!\!&\simeq\!\!\!&}
\newcommand{\defa}{\!\!\!&\equiv\!\!\!&}

\newcommand{\simgt}{\stackrel{>}{{}_\sim}}
\newcommand{\simlt}{\stackrel{<}{{}_\sim}}

\newcommand{\tl}[1]{\tilde{#1}}

\newcommand{\tr}{{\rm tr}}

\newcommand{\der}{\partial}
\newcommand{\dr}{\!\!d}
\newcommand{\hc}{{\rm h.c.}}
\newcommand{\ie}{{i.e.}}

\newcommand{\vev}[1]{\langle #1 \rangle}
\newcommand{\Lvev}[1]{\left\langle #1 \right\rangle}
\newcommand{\brkt}[1]{\left( #1 \right)}
\newcommand{\brc}[1]{\left\{ #1 \right\}}
\newcommand{\sbk}[1]{\left[ #1 \right]}
\newcommand{\abs}[1]{\left| #1 \right|}

\renewcommand{\Re}{{\rm Re}\,}
\renewcommand{\Im}{{\rm Im}\,}

\newcommand{\cF}{{\cal F}}

\newcommand{\cL}{{\cal L}}

\newcommand{\cN}{{\cal N}}
\newcommand{\cO}{{\cal O}}

\newcommand{\cQ}{{\cal Q}}
\newcommand{\cR}{{\cal R}}

\newcommand{\cV}{{\cal V}}

\newcommand{\cW}{{\cal W}}

\newcommand{\cY}{{\cal Y}}
\newcommand{\cZ}{{\cal Z}}

\newcommand{\bQ}{{\mathbb Q}}
\newcommand{\bV}{{\mathbb V}}

\newcommand{\Led}{L_{\rm ED}}
\newcommand{\Tb}{T_{\rm b}}
\newcommand{\Ts}{T_{\rm s}}
\newcommand{\taub}{\tau_{\rm b}}
\newcommand{\taus}{\tau_{\rm s}}
\newcommand{\rhob}{\rho_{\rm b}}
\newcommand{\rhos}{\rho_{\rm s}}

\begin{document}
\thispagestyle{empty}

\baselineskip=12pt

%{\small \noindent \mydate \hfill }

\begin{flushright}
KEK-TH-1699 \\
WU-HEP-14-01 
\end{flushright}

\baselineskip=35pt plus 1pt minus 1pt

\vskip 1.5cm

\begin{center}
{\LARGE\bf Natural realization of large extra dimension 
in 5D supersymmetric theory}

\vspace{1.5cm}
\baselineskip=20pt plus 1pt minus 1pt

\normalsize

{\large\bf Yutaka Sakamura}${}^{1,2}\!${\def\thefootnote{\fnsymbol{footnote}}
\footnote[1]{E-mail address: sakamura@post.kek.jp}} 
{\large\bf and Yusuke Yamada}${}^3\!${\def\thefootnote{\fnsymbol{footnote}}
\footnote[2]{E-mail address: yuusuke-yamada@asagi.waseda.jp}}

\vskip 1.0em

${}^1${\small\it KEK Theory Center, Institute of Particle and Nuclear Studies, 
KEK, \\ Tsukuba, Ibaraki 305-0801, Japan} \\ \vspace{1mm}
${}^2${\small\it Department of Particles and Nuclear Physics, \\
The Graduate University for Advanced Studies (Sokendai), \\
Tsukuba, Ibaraki 305-0801, Japan} 

\vskip 1.0em

${}^3${\small\it Department of Physics, Waseda University, \\ 
Tokyo 169-8555, Japan}

\end{center}

\vskip 1.0cm
\baselineskip=20pt plus 1pt minus 1pt

\begin{abstract}
An exponentially large extra dimension can be naturally realized 
by the Casimir energy and the gaugino condensation 
in 5D supersymmetric theory. 
The model does not require any hierarchies among the 5D parameters. 
The key ingredient is an additional modulus other than the radion, 
which generically exists in 5D supergravity. 
SUSY is broken at the vacuum, which can be regarded as 
the Scherk-Schwarz SUSY breaking. 
We also analyze the mass spectrum and discuss some phenomenological aspects.  
\end{abstract}

%%%%%%%%%%%%%%%%%%%%%%%%
\newpage

\section{Introduction}
Models with extra dimensions are intriguing candidates for the new physics, 
and have been investigated in a vast number of articles. 
%A large number of scenarios with various magnitudes of the compactification scales
%have been proposed so far. 
%In every model, the size of the extra dimensions are assumed to be 
%much larger than the Planck length, which is the fundamental length of the gravity. 
Such models were originally proposed in order to explain the large hierarchy 
between the electroweak scale~$M_{\rm EM}$ 
and the Planck scale~$M_{\rm Pl}$. 
In the warped spacetime~\cite{Randall:1999ee}, 
for example, $M_{\rm EM}$ emerges only from the five-dimensional (5D) parameters 
that are roughly of the same order as $M_{\rm Pl}$. 
In the flat spacetime, however, the smallness of $M_{\rm EM}/M_{\rm Pl}$ is
just translated into largeness of the size of the extra dimensions~$\Led$   
compared to the Planck 
length~$M_{\rm Pl}^{-1}$~\cite{Antoniadis:1990ew,ArkaniHamed:1998rs,Antoniadis:1998ig}. 
We should note that $\Led$ is not a parameter of a theory 
but a quantity dynamically determined by 
some moduli stabilization mechanism. 
In contrast to models in the warped geometry~\cite{Goldberger:1999uk}, 
most of known models on the flat spacetime need to admit 
some hierarchies among the fundamental parameters 
in order to realize the tiny ratio~$M_{\rm EM}/M_{\rm Pl}$. 

In our previous work~\cite{Sakamura:2013rba}, we have shown  
that an exponentially large extra dimension can be obtained 
without introducing any hierarchies among the model parameters 
in the context of 5D supergravity. 
The moduli is stabilized by the Casimir energy and 
a superpotential term induced by the gaugino condensation. 
The purpose of this paper is to investigate this scenario further 
in more general setup, 
and understand the essential structure of the model 
by comparing a model with a similar setup 
in Ref.~\cite{vonGersdorff:2003rq} that does not generate a large extra dimension. 
We also discuss the phenomenological aspects of the model 
by evaluating the mass spectrum of the moduli and the superparticles. 
The latter spectrum largely depends on whether the gauge and matter fields 
are in the bulk or on the boundary. 

The paper is organized as follows. 
In Sec.~\ref{Casimir_stb}, we provide a compact review of 
the stabilization of the extra dimension by the Casimir effect 
in a simple setup. 
In Sec.~\ref{LED}, we extend the model in Sec.~\ref{Casimir_stb} 
by introducing an additional modulus, and show that an exponentially large 
extra dimension is naturally realized. 
In Sec.~\ref{mass_spectrum}, the mass spectrum of our model is discussed 
in various cases according to where the gauge and matter fields live. 
Sec.~\ref{summary} is devoted to the summary. 
In Appendix~\ref{SSbreaking}, we show the consistency of 
our formula for the effective potential 
with the previous works~\cite{vonGersdorff:2003rq,vonGersdorff:2002tj} 
by noting the equivalence of 
supersymmetry (SUSY) breaking by a boundary constant superpotential 
and by the Scherk-Schwarz twisted boundary condition~\cite{Scherk:1978ta}.

\section{Radion stabilization by Casimir energy} \label{Casimir_stb}
In this section, we provide a brief review of 
the radius stabilization by the Casimir energy~\cite{vonGersdorff:2003rq,Ponton:2001hq} 
in the superfield formalism. 
%Since the Scherk-Schwarz twisted boundary condition is not consistent 
%with the warped geometry~\cite{Hall:2003yc,Abe:2005wn}, 

We consider a 5D flat spacetime compactified on $S^1/Z_2$ 
whose fundamental region is denoted as $0\leq y\leq L$. 
The physical field content consists of hypermultiplets~$\bQ_a$, 
where the index~$a$ labels the gauge multiplets,  
vector multiplets~$\bV_G$ ($G=SU(3)_C,SU(2)_L,U(1)_Y,\cdots$), 
and the gravitational multiplet. 
These multiplets can be expressed in terms of $N=1$ 
superfields~\cite{ArkaniHamed:2001tb,Paccetti:2004ri,Abe:2004ar} as
$\bQ_a=(\cQ_a,\cQ_a^c)$ and $\bV_G=(\cV_G,\Sgm_G)$, 
where $\cV_G$ is an $N=1$ vector superfield and the others are chiral superfields. 
The matter superfields~$Q_a$ and the gauge superfields~$V_G$ 
in the SUSY standard model 
can be identified with the zero-modes of $\cQ_a$ and $\cV_G$ whose $Z_2$-parities 
are even at both boundaries. 
In addition to these, there may exist antiperiodic fields 
that have opposite $Z_2$-parities at $y=0$ and $y=L$. 
They do not have zero-modes, but contribute to the Casimir energy. 
We introduce $n_a^P$ periodic and $n_a^A$ antiperiodic hypermultiplets~$\bQ_a$, 
and $n_V^P$ periodic and $n_V^A$ antiperiodic vector multiplets. 

In the 4D effective theory, there also appears a chiral superfield~$T$ 
that comes from the gravitational multiplet, which is called the radion superfield. 
Its lowest component $T|_0$ is identified as
\be
 T|_0 \equiv \tau+i\rho = \Led\abs{\vph_C}-i\rho, \label{def:T|_0}
\ee
where $\Led\equiv\int_0^Ldy\;e_y^{\;\;4}$ is the size of 
the extra dimension, $\rho$ is the Wilson line phase 
for the graviphoton field along the extra dimension, 
and $\vph_C$ is the compensator scalar, 
which will be fixed by the superconformal gauge-fixing condition explained later. 

The 5D bulk Lagrangian is expressed in terms of 
the $N=1$ superfields~\cite{Paccetti:2004ri,Abe:2004ar}. 
In addition, we introduce the following boundary superpotential 
terms localized at $y=0$. 
\be
 W^{(0)} = W_0+W_{\rm yukawa}(\cQ), 
\ee
where $W_0$ is a constant and $W_{\rm yukawa}$ 
contains the Yukawa couplings. 
We will not introduce the K\"ahler potentials nor the gauge kinetic functions 
at the boundaries until Sec.~\ref{mass_spectrum}. 
 
The 4D effective Lagrangian is expressed as~\footnote{
We have dropped the gravitational fluctuation modes. 
Their dependence on the superspace is provided in Ref.~\cite{Sakamura:2011df}. 
} 
\be
 \cL = -\sbk{\int\dr^2\tht\;\sum_G\frac{f_G}{2}\tr\brkt{\cW_G^2}+\hc}
 +\int\dr^4\tht\;\abs{\phi_C}^2\Omg
 +\sbk{\int\dr^2\tht\;\phi_C^3W+\hc}, \label{Leff}
\ee
where $\phi_C$ is the chiral compensator superfield whose lowest component is $\vph_C$,
and $\cW_G^\alp$ is the field strength superfield for $V_G$. 
The holomorphic functions~$f_G$ and $W$ 
are the gauge kinetic functions and the superpotential, 
which are derived as~\cite{Luty:2000ec}
\bea
 f_G(T) \eql C_G T, \;\;\;\;\;
 W(Q,T) = W_0+W_{\rm yukawa}(Q), 
\eea
where $C_G$ is a constant. 
Notice that there is no bulk contribution to $W$ because $N=2$ SUSY in the bulk 
prohibits it. 
The real function~$\Omg$ is related to the K\"ahler potential~$K$ by 
$\Omg=-3e^{-K/3}$. 

In the absence of other moduli than $T$, 
the $T$-dependent part of $\Omg$ is given by~\footnote{
General formula for the one-loop correction to $\Omg$ is calculated 
in Ref.~\cite{Sakamura:2013wia}, where only periodic fields are considered.  
Contribution of the antiperiodic fields is obtained from the formula by 
taking a limit of infinite brane mass terms at one of the boundaries. 
}  
\be
 \Omg = -3\Re T-\frac{\xi(\Re T)}{(\Re T)^2}+\cdots,  \label{expr:Omg}
\ee
where the ellipsis denotes terms independent of $T$, and 
\bea
 \xi(\tau) \defa \frac{\zeta(3)}{32\pi^2}\left\{\sum_a n_a^P\cZ_P\brkt{c_a\tau}
 -\frac{3}{4}\sum_a n_a^A\cZ_A\brkt{c_a\tau} 
 -n_V^P+\frac{3}{4}n_V^A-2\right\}, \nonumber\\
  \cZ_P(x) \defa -\frac{4}{\zeta(3)}\int_{\abs{x}}^\infty\dr\lmd\;\lmd
 \ln\brkt{1-e^{-2\lmd}}, \nonumber\\
 \cZ_A(x) \defa \frac{16}{3\zeta(3)}\int_{\abs{x}}^\infty\dr\lmd\;
 \lmd\ln\brkt{1+\frac{\lmd-x}{\lmd+x}e^{-2\lmd}}.  \label{def:xicZ}
\eea
The first term in (\ref{expr:Omg}) is the well-known radion K\"ahler potential 
at tree level. 
It has the no-scale structure, and does not induce the potential for the radion. 
This structure is broken by the second term in (\ref{expr:Omg}), 
which is the one-loop correction. 
%The terms proportional to $\xi_V$ and $\xi_a$ are the contributions 
%from the loop of the vector and the gravitational multiplets 
%and from the hypermultiplets, respectively. 
A constant~$c_a$ in the arguments of $\cZ_{P,A}$ is defined as $c_a\equiv \frac{M_a}{M_5}$, 
where $M_a$ is a bulk mass for $\bQ_a$ and $M_5$ is the 5D Planck mass,\footnote{ 
In SUGRA, every mass parameter is introduced by gauging some isometry. 
The bulk mass~$M_a$ is associated with the gauging of the $U(1)$ symmetry 
that rotates the phases of $\cQ_a$ and $\cQ_a^c$ with opposite charges. 
The ratio~$c_a$ is the gauge coupling constant for this gauging. 
}
and $\zeta(s)$ is the Riemann's zeta function. 
Functions~$\cZ_P(x)$ and $\cZ_A(x)$ are normalized so that $\cZ_P(0)=\cZ_A(0)=1$. 
The bulk mass~$M_a$ for $\bQ_a$ (or $c_a$) controls the wave function profile 
along the extra dimension, and the flip~$c_a\to -c_a$ changes the boundary 
toward which the wave function localizes. 
This is the reason why $\cZ_P(x)$ is symmetric while $\cZ_A(x)$ is not  
under $x\to -x$. 
Since both functions exponentially decreases in the region~$\abs{x}\simgt 1$, 
only modes spread over the bulk contribute to $\Omg$ in the one-loop diagrams. 
There is no brane-to-brane contributions to $\Omg$~\cite{Rattazzi:2003rj,Gregoire:2004nn}, 
which correspond to the third term in (4.1) of Ref.~\cite{Sakamura:2013rba}, 
because we introduce the matter-dependent superpotential only at $y=0$. 

As shown in Ref.~\cite{vonGersdorff:2003rq}, the extra dimension is stabilized 
by this setup. 
(See Appendix~\ref{SSbreaking}.)
For later convenience, however, we introduce a nonabelian gauge sector in the bulk, 
in which the gaugino condensation occurs. 
Since the gauge kinetic function is proportional to $T$, 
the following superpotential term is induced.  
\be
 W = A e^{-aT}+\cdots, 
\ee
where $A$ is a complex constant and $a$ is a real constant of $\cO(4\pi^2)$. 

Then the effective potential~$V_{\rm eff}$ is calculated as 
\bea
 V_{\rm eff} \eql \abs{\vph_C}^4\brc{\frac{\abs{W_0}^2\cF(\tau)}{\tau^4}
 +4a\abs{W_0A}e^{-a\tau}\cos(a\rho-\vth)}+\cdots,  \label{Veff:single0}
\eea
where $\vth\equiv\arg(\bar{W}_0A)$, and 
\be
 \cF(\tau) \equiv 6\xi(\tau)-4\tau\xi'(\tau)+\tau^2\xi''(\tau). 
\ee
We have dropped higher-order terms in the loop factor~$\zeta(3)/32\pi^2$ or $e^{-a\tau}$ 
and $Q_a$-dependent terms. 

In order to obtain the ordinary Poincar\'{e} SUGRA, we have to 
impose the superconformal gauge-fixing conditions. 
According to the action formula in Ref.~\cite{Kugo:1982cu,Kugo:2002vc}, 
the gravitational part of (\ref{Leff}) is 
\be
 \cL_{\rm grav} = \frac{\sqrt{-g}}{6}\abs{\vph_C}^2\Omg|_0\cR+\cdots, 
\ee
where $g\equiv\det(g_{\mu\nu})$, $\cR$ is the Ricci scalar, 
and $\Omg|_0$ is the lowest component of $\Omg$.  
Thus, the condition to obtain the canonically normalized Einstein term is 
\be
 \vph_C = \brkt{-\frac{3}{\Omg|_0}}^{1/2} 
 = \tau^{-\frac{1}{2}}+\cdots,  \label{D-fixing}
\ee
where we have used (\ref{expr:Omg}) at the second equality, 
and the ellipsis denotes terms suppressed by $\zeta(3)/32\pi^2$
and $Q_a$-dependent terms. 
Throughout this paper, we take the unit of 4D Planck mass, \ie, $M_{\rm Pl}=1$. 
In this gauge, we find from (\ref{def:T|_0}) that~\footnote{
Note that $\tau=\Led+\cdots$ in the unit of $M_5$ 
since $M_5=\Led^{-1/3}$. 
}
\be
 \tau = \Led^{2/3}+\cdots.  \label{ReT-Lphys}
\ee

From (\ref{Veff:single0}) with (\ref{D-fixing}), 
the minimization conditions for $V_{\rm eff}$ are 
\bea
 &&-6\cF(\tau)+\tau\cF'(\tau)+4a(a\tau+2)\abs{\frac{A}{W_0}}\tau^4 e^{-a\tau} = 0, 
 \nonumber\\
 &&\cos(a\rho-\vth) = -1.  \label{minimization:single}
\eea
When the 5D masses for $\cQ_a$ are zero (\ie, $c_a=0$), 
$\xi(\tau)$ becomes constant and 
the first equation is reduced to
\be
 (a\tau+2)\tau^4 e^{-a\tau} = \frac{3\xi}{2a}\abs{\frac{W_0}{A}}. 
\ee
This does not have a solution in the region~$\tau\gg 1$ 
unless $\abs{W_0/A}$ is exponentially small.\footnote{
A solution to (\ref{minimization:single}) is generically a (SUSY-breaking) 
anti de Sitter (AdS) vacuum. 
So we need to uplift the vacuum energy to achieve the 4D Minkowski space 
by introducing additional sector. 
Such uplifting sector affects the vacuum solution, 
%from that of (\ref{minimization:single}), 
but this does not improve the situation drastically. }

In Ref.~\cite{vonGersdorff:2003rq}, the extra dimension is stabilized 
by the $\tau$-dependence of $\cF(\tau)$ 
in the absence of the gaugino condensation ($A=0$). 
In this case, the $\cO(\mbox{TeV})$ Kaluza-Klein (KK) scale is obtained by assuming 
the bulk hypermultiplet mass~$M_H$ is also of $\cO({\rm TeV})$. 
In any case, we have to admit a large hierarchy among the fundamental scales 
of the 5D theory. 
(See (\ref{rel:5Dparameters}) in Appendix~\ref{SSbreaking}.)
This stems from the assumption that there is only one modulus, \ie, the radion. 
Hence we will extend the model in the next section 
so that an additional modulus appears in the effective theory.

\section{Realization of large extra dimension} \label{LED}
In this section, we extend the model in the previous section, 
and construct a model in which the extra dimension is stabilized 
at an exponentially large size without introducing any hierarchical parameters 
in 5D theory. 

Notice that the vacuum expectation value (VEV) 
of the radion~$\tau$ determines both 
the size of the extra dimension~$\Led$ and 
the gauge coupling constant of the condensation sector. 
The former is relevant to the volume suppression of the Casimir energy~$\tau^{-6}$, 
and the latter is to the exponential factor in the second term 
in (\ref{Veff:single0}).\footnote{
The radion also determines the wave function profile of $Q_a$. 
}
In a case where additional moduli appear in the 4D effective theory, 
the above two quantities can be controlled by different moduli separately. 
Such additional moduli originate from 5D vector multiplets 
whose 4D vector components are $Z_2$-odd at both boundaries. 
They commonly exist if the 5D theory is an effective theory of 
a higher dimensional theory, such as 10D superstring theory. 

In the following, we consider a case that there is one additional modulus 
other than the radion. 
These moduli generically mix with each other, 
so it is convenient to treat them on equal footing 
by denoting them as $T^I$ ($I=1,2$).  
The mixing is described 
by the cubic polynomial called the norm function~$\cN(\tau)$, 
\be
 \cN(\tau) = \tl{C}_0(\tau^1)^3+3\tl{C}_1(\tau^1)^2\tau^2
 +3\tl{C}_2\tau^1(\tau^2)^2+\tl{C}_3(\tau^2)^3, 
\ee
where $\tl{C}_{0,1,2,3}$ are real constants, 
and $\tau^I\equiv\Re T^I$. 
This corresponds to the prepotential of 4D $N=2$ SUSY theory. 
Then (\ref{ReT-Lphys}) is extended to 
\be
 \cN(\tau) = \Led^2+\cdots. 
\ee
Hence we are interested in a situation where $\cN(\vev{\tau})\gg 1$. 
It is convenient to rotate the moduli fields~$(T^1,T^2)$ to 
a new basis~$(\Tb,\Ts)$ so that the norm function takes the following form. 
\be
 \cN(\tau) = \taub^3+3C_1\taub^2\taus+3C_1^2\taub\taus^2+C_3\taus^3, 
 \label{redef:cN}
\ee
where $C_1$ and $C_3$ are real constants.\footnote{
The model discussed in our previous work~\cite{Sakamura:2013rba} 
is the case of $C_1=0$ and $C_3<0$. 
}  
Note that this redefinition is always possible. 
We will look for a vacuum where $\vev{\taub}\gg\vev{\taus}=\cO(1)$. 
In such a vacuum, $\taub$ is almost identified with the radion. 

Before discussing the moduli stabilization, we comment on 
the 4D coupling constants in the effective theory. 
The gauge coupling constants~$g_G$ are read off from the gauge kinetic functions as 
\be
 g_G = f_G(\vev{\tau})^{-\frac{1}{2}} 
 = (C_G^{\rm b}\vev{\taub}+C_G^{\rm s}\vev{\taus})^{-\frac{1}{2}}, 
\ee
where $C_G^{\rm b}$ and $C_G^{\rm s}$ are real constants. 
Thus we assume that all the gauge kinetic functions depend only on $\Ts$, 
\ie, $C_G^{\rm b}=0$. 
Otherwise, the 4D gauge couplings become too small 
by the large VEV of $\taub$.\footnote{
Notice that the 5D gauge coupling constants~$g^{(5)}_G=g_G \Led^{1/2}$ 
are not the fundamental parameters in 5D SUGRA, 
but are determined by VEVs of the moduli. 
They are exponentially large compared to the 5D Planck mass~$M_5=\Led^{-1/3}$ 
in our case. 
} 

The Yukawa coupling constants~$y_{abc}$ are read off from the matter part of 
$\Omg$~\cite{Abe:2006eg,Abe:2008an,Abe:2011rg}, 
\be
 \Omg^{\rm matter} = 2\cN^{1/3}(\Re T)\sum_a Y_{c_a}(\Re T)\abs{Q_a}^2+\cdots, 
 \label{def:Omg^matter}
\ee
where
\be
 Y_c(\Re T) \equiv \frac{1-e^{-2c\cdot\Re T}}{2c\cdot\Re T}.  \label{def:Y_c}
\ee
After the canonical normalization of $Q_a$, we obtain 
\be
 y_{abc} = \frac{\lmd_{abc}}{\sqrt{
 \vev{8\cN(\tau)Y_{c_a}(\tau)Y_{c_b}(\tau)Y_{c_c}(\tau)}}}, 
\ee
where $\lmd_{abc}$ are the coupling constants in 
the boundary superpotential term~$W_{\rm yukawa}(\cQ)$. 
Since $\cN(\vev{\tau})$ is exponentially large in our case, 
the standard model matter fields~$Q_a$ must be charged for 
the radion multiplet~$\bV_{\rm b}$ with positive charges, \ie, $c_a^{\rm b}>0$, 
in order to obtain the realistic values of the Yukawa couplings.\footnote{
We cannot explain the fermion mass hierarchy by the wave function localization 
in this case. 
We need some mechanism that generates the hierarchical structure of $\lmd_{abc}$. 
} 
Namely, the wave functions of $Q_a$ strongly localize toward $y=0$ 
where $W_{\rm yukawa}(\cQ)$ exists. 
Such localized modes do not contribute to the Casimir energy 
as mentioned in the previous section. 

The effective theory has the following superpotential. 
\be
 W = W_0+A e^{-aT_s}+W_{\rm yukawa}(Q_a), 
\ee
where $W_0$ and $A$ are complex constants and $a=\cO(4\pi)$ is a real constant.  
Notice that the second term induced by the gaugino condensation is independent of $\Tb$ 
since we have assumed that the gauge kinetic functions 
only depend on $\Ts$. 

Now $\Omg$ in (\ref{expr:Omg}) is modified as
\be
 \Omg = -3\cN^{1/3}(\Re T)
 -\frac{\xi(\Re T)}{\cN^{2/3}(\Re T)}+\cdots.  \label{multi:Omg}
\ee
Here $c_a\tau$ in the definition of $\xi(\tau)$ in (\ref{def:xicZ}) 
is now understood as $c_a\cdot\tau\equiv c_a^{\rm b}\taub+c_a^{\rm s}\taus$. 
Let us consider a case that all the standard model fields are in the bulk 
and no other zero-modes exist, 
\ie, $n_V^P=12$ and $\sum_an_a^P=52$. 
We should note that $\cZ_P(c_a\tau)\simeq 0$ 
because the matter fields strongly localize toward the boundary. 
By the same reason, we can neglect contributions of antiperiodic hypermultiplets 
that are charged for $\bV_b$. 
Therefore, $\xi(\tau)$ is expressed as 
\be
 \xi(\tau) = \frac{3\zeta(3)}{128\pi^2}\brc{
 n_V^A-\sum_a n_a^A\cZ_A\brkt{c_a^{\rm s}\taus}-20}. \label{simple:xi}
\ee
Notice that there is additional contribution~$-\zeta(3)/32\pi^2$ 
compared to (\ref{def:xicZ}), which comes from the additional modulus multiplet. 
In the following, we focus on a case that $n_a^A=0$ to simplify the discussion. 
Then $\xi(\tau)$ becomes a constant~$\xi_0\equiv 3\zeta(3)(n_V^A-20)/128\pi^2$. 
The effective potential for the moduli is now calculated as
\bea
 V_{\rm eff} \eql \abs{\vph_C}^4e^{K/3}
 \brc{\abs{W}^2\brkt{K_IK^{I\bar{J}}K_{\bar{J}}-3}
 +\brkt{K^{I\bar{J}}K_{\bar{J}}\bar{W}W_I+\hc}
 +K^{I\bar{J}}W_I\bar{W}_{\bar{J}}} \nonumber\\
 \eql \frac{6\xi_0\abs{W_0}^2}{\taub^6}
 +\frac{4a\taus\abs{W_0A}e^{-a\taus}}{\taub^3}\cos(a\rhos-\vth)
 +\frac{2a^2\abs{A}^2e^{-2a\taus}}{3(C_1^3-C_3)\taus}
 +\cdots, \label{expr:Veff2}
\eea
where $\vth\equiv\arg(\bar{W}_0A)$, and the ellipsis denotes terms suppressed 
by $1/\taub$. 
We have used at the second equality that
\be
 \abs{\vph_C}^2 = -\frac{3}{\Omg} = \frac{1}{\taub}+\cO(\taub^{-2}). 
\ee
The third term in (\ref{expr:Veff2}), which was dropped in (\ref{Veff:single0}), 
is now crucial to stabilize $\taus$. 

From the minimization condition for $V_{\rm eff}$, we obtain 
\bea
 &&\taub^3 = \frac{3\xi_0}{a\taus}\abs{\frac{W_0}{A}}e^{a\taus}, 
 \nonumber\\
 &&2(a\taus-1)
 -\frac{\xi_0(2a\taus+1)}{(C_1^3-C_3)\taus^3} = 0, \nonumber\\
 &&\cos(a\rhos-\vth) = -1.  \label{min_cond}
\eea
Since $a=\cO(4\pi^2)\gg\cO(1)$, the second equation is solved as
\be
 \taus = \brkt{\frac{\xi_0}{C_1^3-C_3}}^{1/3}+\cO(a^{-1}). \label{VEV:tau_s}
\ee
If this value is of $\cO(1)$, we obtain an exponentially large extra dimension. 
For example, $\Led\simeq 10^{15},10^7,10^3$ 
for $(n_V^A,\abs{W_0/A},a,C_1^3-C_3)=(50,1,8\pi^2,0.1)$, 
$(40,1,8\pi^2,0.5)$, $(40,1,4\pi^2,0.5)$, respectively.  

We have assumed that the gauge kinetic function of the condensation sector~$f_C(T)$  
is independent of $\Tb$. 
Even if this is not satisfied, we can always redefine the moduli so that $f_C(T)=\Ts$. 
However, this redefinition breaks the structure of the norm function in (\ref{redef:cN}). 
In such a case, the third term in (\ref{expr:Veff2}) is suppressed by $\taub$ 
instead of $\taus$, and thus is negligible. 
Then we need the $\taus$-dependence of $\xi(\tau)$ in (\ref{simple:xi}) 
by considering a case that $n_a^A\neq 0$, 
in order for $\taus$ to be stabilized at an $\cO(1)$ value. 
We should also note that a vacuum with an exponentially large $\vev{\taub}$ exists 
even if $W$ has another gaugino condensation terms that depend on $\Tb$ 
because such terms are highly suppressed around the vacuum.

\section{Mass spectrum} \label{mass_spectrum}
\subsection{Uplifting and moduli masses}
The solution to (\ref{min_cond}) is an AdS vacuum with 
a negative cosmological constant: 
\be
 V_{\rm eff}(\vev{\tau}) = -\frac{18\xi_0\abs{W_0}^2}
 {(2a\vev{\taus}+1)\vev{\taub^6}} < 0.  \label{NCC}
\ee
In order to achieve the 4D Minkowski spacetime, we cancel this with 
a nonvanishing F-term of a chiral superfield~$X$ in 4D effective theory, 
which is tuned as 
\be
 \abs{F^X}^2 = K_{X\bar{X}}^{-1}\abs{V_{\rm eff}(\vev{\tau})}, \label{mag:F^X}
\ee
where $K_{X\bar{X}}$ is the $(X,\bar{X})$-component of the K\"ahler metric. 
Since the negative cosmological constant~(\ref{NCC}) is exponentially small, 
the effects of the uplifting are tiny. 
However, as we will show in Sec.~\ref{softmass}, 
it provides nonnegligible contributions to the superparticle masses 
in some cases. 

After the canonical normalization, we obtain the mass matrix for the moduli. 
The radion~$\taub$ and the nongeometric modulus~$\taus$ generically 
have a mixing and their mass squared matrix is 
\be
 M^2_\tau \equiv \begin{pmatrix} \sqrt{\frac{2}{K_1}} & 0 \\
 0 & \sqrt{\frac{2}{K_2}} \end{pmatrix}U_K
 \begin{pmatrix} \frac{\der^2V_{\rm eff}}{\der\taub^2} & 
 \frac{\der^2V_{\rm eff}}{\der\taub\der\taus} \\
 \frac{\der^2V_{\rm eff}}{\der\taub\der\taus} & 
 \frac{\der^2V_{\rm eff}}{\der\taus^2} \end{pmatrix} U_K^{-1}
 \begin{pmatrix} \sqrt{\frac{2}{K_1}} & 0 \\
 0 & \sqrt{\frac{2}{K_2}} \end{pmatrix},  \label{mass_matrix}
\ee
where $K_1$, $K_2$, and $U_K$ are the eigenvalues and the diagonalizing matrix 
of the K\"ahler metric, and given by 
\bea
 K_1 \eql \frac{3(1+C_1^2)}{4\taub^2}+\cdots, \;\;\;\;\;
 K_2 = \frac{3(C_1^3-C_3)\taus}{2(1+C_1^2)\taub^3}+\cdots, \nonumber\\
 U_K \eql \frac{1}{\sqrt{1+C_1^2}}\begin{pmatrix} 1-C_1^2\dlt & C_1(1+\dlt) \\
 -C_1(1+\dlt) & 1-C_1^2\dlt \end{pmatrix}+\cdots, \;\;\;\;\;
 \dlt \equiv \frac{2(C_1^3-C_3)\taus}{(1+C_1^2)^2\taub}. 
\eea
where the ellipses are terms suppressed by $\taus/\taub$, 
and (\ref{mass_matrix}) is evaluated at the vacuum. 
By diagonalizing (\ref{mass_matrix}), we obtain the moduli masses as follows. 
\bea
 m_{\tau_{\rm l}} \sma \begin{cases} \frac{12\sqrt{6\xi_0\vev{\taus}}}{\sqrt{a}C_1}
 \frac{\abs{W_0}}{\vev{\taub^3}}, & (C_1 \neq 0) \\
 \frac{12a\sqrt{6\xi_0}}{\sqrt{a\vev{\taus}}}\frac{\abs{W_0}}{\vev{\taub^3}}, 
 & (C_1 = 0) \end{cases} \;\;\;\;\;\;\;
 m_{\rhob} \simeq 0, \nonumber\\
 m_{\tau_{\rm h}} \sma m_{\rhos} 
 \simeq \begin{cases} \frac{4C_1a\vev{\taus}\abs{W_0}}{\vev{\taub^{3/2}}}, 
 & (C_1 \neq 0) \\
 \frac{4a\vev{\taus}\abs{W_0}}{\vev{\taub^{3/2}}},  & (C_1 = 0) \end{cases} 
 \label{m_tau}
\eea
where $\tau_{\rm l}$ ($\tau_{\rm h}$) is the lighter (heavier) mass eigenstate. 
When $C_1=0$, $(\tau_{\rm l},\tau_{\rm h})$ are almost reduced to
$(\taub,\taus)$ up to the normalization factors.

\subsection{SUSY-breaking F-terms}
The vacuum~(\ref{min_cond}) breaks SUSY because of the constant superpotential~$W_0$, 
which is equivalent to the Scherk-Schwarz twisted boundary condition. 
(See Appendix~\ref{SSbreaking}.)
The gravitino mass is then 
\be
 m_{3/2} = \vev{e^{K/2}W} \simeq \frac{\abs{W_0}}{\vev{\taub^{3/2}}} 
 \simeq \frac{\abs{W_0}}{\Led}. 
\ee
The moduli F-terms are estimated from the equations of motion as 
\bea
 \frac{F^{\Tb}}{2\taub} \eql \frac{\bar{W}_0}{\vev{\taub^{3/2}}}
 \brc{1+\cO(\vev{\taub^{-1}})}
 \simeq m_{3/2}, \nonumber\\
 \frac{F^{\Ts}}{2\taus} \eql \frac{\bar{W}_0}{\vev{a\taus\taub^{3/2}}}
 \frac{\xi_0+2(C_1^3-C_3)\vev{\taus^3}}{2(C_1^3-C_3)\vev{\taus^3}}
 \brc{1+\cO(\vev{\taub^{-1}})}
 \simeq \frac{m_{3/2}}{a\vev{\taus}}. 
\eea
The compensator F-term~$F^{\phi_C}$ is negligible 
as a result of the (approximate) no-scale structure. 

The uplifting superfield~$X$ can originate from either a bulk hypermultiplet 
or a brane-localized chiral multiplet. 
Since the K\"ahler potential in each case is given by 
\be
 \Omg = \begin{cases} -\cN^{1/3}(\Re T)\brc{3-2Y_{c_X}(\Re T)\abs{X}^2+\cdots}, 
 & \mbox{(Bulk origin)} \\
 -3\cN^{1/3}(\Re T)+h_X\abs{X}^2+\cdots, & \mbox{(Brane origin)} \end{cases}
\ee
where $Y_c(\tau)$ is defined in (\ref{def:Y_c}), $(c_X^{\rm b},c_X^{\rm s})$ 
are charges of $X$ for $(\bV_{\rm b},\bV_{\rm s})$, 
and $h_X$ is a real constant,  
the K\"ahler metric is 
\be
 K_{X\bar{X}} = -3\brkt{\frac{\Omg_{X\bar{X}}}{\Omg}-\frac{\abs{\Omg_X}^2}{\Omg^2}} 
 \simeq \begin{cases} 2Y_{c_X}(\tau), & (\mbox{Bulk origin}) \\
 \frac{h_X}{\cN^{1/3}(\tau)}, & (\mbox{Brane origin}) \end{cases}
\ee
where we have assumed that $\abs{X}\ll 1$. 
Hence the F-term of $X$ is estimated from (\ref{NCC}) and (\ref{mag:F^X}) as 
\bea
 \abs{F^X} \eql \begin{cases} \frac{3m_{3/2}}{\vev{\taub^{3/2}}}
 \sqrt{\frac{\xi_0}{(2a\vev{\taus}+1)Y_{c_X}(\vev{\tau})}}, 
 & (\mbox{Bulk origin}) \\
 \frac{3m_{3/2}}{\vev{\taub}}\sqrt{\frac{2\xi_0}{(2a\vev{\taus}+1)h_X}}. 
 & (\mbox{Brane origin}) \end{cases}
\eea
When $X$ lives in the bulk, 
$F^X$ is negligible for $c_X\cdot\vev{\tau}<0$, 
and it grows up to the same order as that in the case of $X$ on the the boundary 
for $c_X\cdot\vev{\tau}>0$. 
Thus we assume that $c_X\cdot\vev{\tau}=0$, 
\ie, $Y_{c_X}(\tau)=1$ in the following. 
In this case, $X$ also contributes to (\ref{simple:xi}) 
and $\xi_0$ in (\ref{expr:Veff2}) is modified as $\xi_0=\zeta(3)(3n_V^A-56)/128\pi^2$. 

The F-terms of the other chiral superfields are negligible. 
Therefore the dominant source of SUSY breaking is $F^{\Tb}$.

\subsection{Superparticle masses} \label{softmass}
In this subsection, we estimate the mass spectrum of the superparticles 
in three cases classified according to where the gauge and the matter fields live. 

\subsubsection{Gauge and matter fields in the bulk} \label{BBcase}
First we discuss a case that both the gauge and the chiral matter fields live 
in the bulk. 
In this case, the gaugino masses~$M_G$ ($G=SU(3)_C,SU(2)_L,U(1)_Y$) are calculated as 
\be
 M_G = \vev{F^I\der_I\ln(\Re f_G(T))} 
 \simeq \Lvev{\frac{F^{T_s}}{2\taus}} \simeq \frac{m_{3/2}}{a\vev{\taus}}. 
 \label{M_G:BB}
\ee
Notice that there is no contribution from $F^{\Tb}$ 
since $f_G(T)$ does not depend on $\Tb$ by assumption. 

The soft scalar masses of $Q_a$ are 
\bea
 m_{Q_a}^2 \eql -F^I\bar{F}^{\bar{J}}\der_I\der_{\bar{J}}
 \ln\brkt{\der_{Q_a}\der_{\bar{Q}_a}\Omg} 
 \nonumber\\
 \sma m_{3/2}^2\brc{1-(c_a\cdot\vev{\tau})^2\cY_a(c_a\cdot\vev{\tau})}, 
\eea
where we have used (\ref{def:Omg^matter}), and 
\be
 \cY(x) \equiv \frac{1+e^{4x}-2e^{2x}(1+2x^2)}{(1-e^{2x})^2x^2}, 
\ee
is even and monotonically decreasing function of $\abs{x}$ and $\cY(0)=1/3$. 
Recall that the quark and lepton superfields are strongly localized 
toward $y=0$, \ie, $c_a\cdot\vev{\tau}\gg 1$, to achieve the observed fermion masses. 
Thus these masses become much smaller than $m_{3/2}$ 
since $\lim_{x\to\infty}x^2\cY(x)=1$. 
This can be understood because such fields are almost regarded as 
the boundary-localized fields, which do not have couplings with the moduli 
at tree level. 
Thus we have to consider the next-leading contributions,  
and obtain 
\be
 m_{Q_a}^2 \simeq \frac{2c_a^{\rm s}\vev{\taus}}{c_a^{\rm b}\vev{\taub}}\cdot m_{3/2}^2
 \ll m_{3/2}^2. 
\ee

Although the scalar masses are much smaller than the gaugino masses 
at the compactification scale~$m_{\rm KK}=\pi/\Led$ in this case, 
the former become comparable to the latter in low energies 
by the renormalization group effect if $m_{\rm KK}$ is high enough, 
for example, $m_{\rm KK}=\cO(10^{16}\mbox{GeV})$. 
This situation is similar to the gaugino mediation~\cite{Chacko:1999mi}, 
and we obtain the flavor universal soft masses in such a case.

\subsubsection{Gauge and matter fields on the boundary}
Next we consider a case that both the gauge fields and 
the chiral matter multiplets live on the boundary~$y=0$. 
Since the brane-localized fields do not couple with the moduli at tree level, 
the contribution from the moduli F-terms does not exist. 
Here we assume the following gauge kinetic function~$f^{(0)}_G$ localized at $y=0$. 
\be
 f^{(0)}_G(X) = k_{0G}+k_{1G} X, 
\ee
where $k_{0G}$ and $k_{1G}$ are $\cO(1)$ constants.\footnote{
Since the R-symmetry is explicitly broken by the constant superpotential~$W_0$, 
there is no reason to forbid the second term. 
}
Then the gaugino masses~$M_G$ are expressed as 
\bea
 M_G \eql \abs{F^I\der_I\ln f_G^{(0)}} = g_G^2k_{1G}\abs{F^X} \nonumber\\
 \eql \begin{cases} \cO\brkt{\frac{g_G^2}{\vev{\taub^{3/2}}}
 \sqrt{\frac{\xi_0}{a\vev{\taus}}}}m_{3/2}, & (\mbox{$X$: Bulk origin}) \\
 \cO\brkt{\frac{g_G^2}{\vev{\taub}}\sqrt{\frac{\xi_0}{a\vev{\taus}}}}m_{3/2}.  
 & (\mbox{$X$: Brane origin}) \end{cases}
\eea
Here the gauge coupling constants are given by 
$g_G^2=(\Re\vev{f_G^{(0)}})^{-1}$. 

As for the chiral matter multiplets, we assume the brane-localized 
K\"ahler potential~$\Omg^{(0)}$ to have a form, 
\be
 \Omg^{(0)} = \sum_a\brkt{h_a\abs{q_a}^2-\kp_{aX}\abs{q_a}^2\abs{X}^2}, 
\ee
where $h_a$ and $\kp_{aX}$ are positive $\cO(1)$ constants, 
and $q_a$ are brane-localized chiral superfields.  
Then the effective K\"ahler potential is calculated as 
\be
 \Omg = \Omg^{(0)}-\frac{\zeta(3)}{8\pi^2\cN(\Re T)}\Omg^{(0)}+\cdots. 
\ee
Therefore, the soft scalar masses for $q_a$ are computed as
\bea
 m_{q_a}^2 \eql -F^I\bar{F}^{\bar{J}}\der_I\der_{\bar{J}}\ln
 \brkt{\der_{q_a}\der_{\bar{q}_a}\Omg} \nonumber\\
 \eql \frac{3\zeta(3)m_{3/2}^2}{2\pi^2\cN(\vev{\tau})}
 +\frac{\kp_{aX}}{h_a}\abs{F^X}^2 
 = \begin{cases} \cO\brkt{\frac{m_{3/2}^2}{\vev{a\taus\taub^3}}}, 
 & (\mbox{$X$: Bulk origin}) \\
 \cO\brkt{\frac{m_{3/2}^2}{\vev{a\taus\taub^2}}}. 
 & (\mbox{$X$: Brane origin}) \end{cases}  \label{m_q:bb}
\eea
The first term in the second line is the contribution of $F^{\Tb}$. 

Hence all the superparticles have masses of the same order of the magnitude, 
which can be set to be $\cO(\mbox{TeV})$. 
In this case, $\Led=\cO(10^7)$ and $m_{3/2}=10^{11}$~GeV 
when $X$ originates from the bulk field, 
and $\Led=\cO(10^8)$ and $m_{3/2}=10^{10}$~GeV 
when $X$ from the brane field.

\subsubsection{Gauge fields in the bulk and matter fields on the boundary}
Finally we consider a case that the gauge fields live in the bulk 
while the matter fields are localized on the boundary~$y=0$. 
In this case, the gaugino masses are obtained by (\ref{M_G:BB}), 
and the soft scalar masses are by (\ref{m_q:bb}). 
Thus the situation is similar to the case in Sec.~\ref{BBcase}, 
and the flavor universal soft masses are obtained 
if $m_{\rm KK}$ is high enough. 

In either case discussed above, 
the higgsino mass can be obtained by adding 
the Giudice-Masiero terms~\cite{Giudice:1988yz} to $\Omg^{(0)}$, 
\be
 \Omg^{(0)}_{\rm GM} = \eta H_uH_d+\hc, \label{GMterm}
\ee
where $H_u$ and $H_d$ are the up- and the down-type Higgs superfields, 
and $\eta$ is a constant. 
Note that these terms cannot be introduced in the bulk 
because of the $N=2$ SUSY structure. 

Because the analysis in this section is performed in 4D effective theory, 
all masses in the above expressions must be laid below $m_{\rm KK}$. 
This condition is satisfied if $\abs{W_0}<(a\vev{\tau_s})^{-1}$. 
However we should note that our mechanism for the realization of 
a large extra dimension still works even when $\abs{W_0}=\cO(1)$, 
although the expressions of $m_{\tau_{\rm h}}$ and $m_{\rhos}$ in (\ref{m_tau}) 
have to be modified.

\subsection{Comment on cosmology}
Before closing this section, we provide some comments on cosmology 
based on our model. 
Notice that the lighter modulus~$\tau_{\rm l}$ and the axionic component~$\rhob$ 
are much lighter than the MSSM superparticles 
in all cases discussed in the previous subsection. 
Such light particles may cause some cosmological problems. 

In the F-term inflation models, $\tau_{\rm l}$ is not stabilized during inflation 
if the Hubble scale at that time~$H_{\rm inf}$ is larger than $m_{3/2}$. 
This so-called the overshooting problem also occurs in models 
in Refs.~\cite{Kachru:2003aw,Balasubramanian:2005zx,Conlon:2005ki}, 
and some solutions to it have been proposed 
in Refs.~\cite{Kallosh:2004yh,Kobayashi:2010rx,Yamada:2012tj}. 
Besides, the radion~$\taub$ generically takes a different value 
from the present minimum during inflation,  
and starts to oscillate after it ends.  
Such oscillation dominates the energy density of the universe,   
%because its lifetime is extremely long due to its tiny gravitational couplings. 
and its decay ruins the successful big bang nucleosynthesis. 
Low-scale inflation may be a way out of these problems. 
In the MSSM inflation model~\cite{Allahverdi:2006iq}, for example, 
$H_{\rm inf}=\cO(0.1\mbox{GeV})$ 
and the correction to the moduli potential during inflation is small. 
So the above problems do not occur. 
However, some fine-tunings among the model parameters are generically required 
to realize low-scale inflation. 

The axionic component~$\rhob$ remains massless and thus contributes 
to the dark radiation. 
It is pointed out in Ref.~\cite{Higaki:2013lra} 
that $\rhob$ produced from the decay of $\tau_{\rm l}$ 
leads to too much dark radiation which contradicts the observation,  
even if $m_{\tau_{\rm l}}\simgt\cO(10\mbox{TeV})$. 
One of the solutions suggested there is 
to increase the partial width of 
the $\tau_{\rm l}$ decay into the standard model particles such as 
the Higgs bosons or the gauge bosons. 
This can be achieved in our model 
by increasing the coupling constant~$\eta$ in (\ref{GMterm}), 
for example.

\section{Summary} \label{summary}
We explicitly showed that an exponentially large extra dimension can be 
naturally realized by the Casimir energy and the gaugino condensation 
in 5D supergravity on $S^1/Z_2$. 
The key ingredient is the nongeometric moduli, 
which are generically present in 5D supergravity. 
The relevant modulus to the Casimir energy is the geometric modulus, 
\ie, the radion. 
However, there is no reason that the same modulus also determines  
the gauge coupling constant of the condensation sector. 
When the relevant modulus to that sector is different 
from the radion, the moduli potential has a minimum 
at an exponentially large VEV for the radion 
even if any hierarchies among the 5D parameters are not assumed. 
Therefore we can dynamically obtain the TeV-scale KK scale 
only from the Planck-scale parameters. 

The potential does not exist at tree level due to the no-scale structure 
of the K\"ahler potential. 
The one-loop correction breaks the structure and generates the potential 
for the moduli. 
The situation is similar to the LARGE volume scenario 
in string theory~\cite{Balasubramanian:2005zx,Conlon:2005ki}, 
but our mechanism does not need any stringy effects and 
works within the field theory. 
Besides we can explicitly calculate the spectrum and effective couplings 
because 5D supergravity is much more tractable than string theory. 

SUSY is broken at the vacuum. 
This is essentially the Scherk-Schwarz SUSY breaking, 
which is equivalent to the constant superpotential 
on the boundary~\cite{vonGersdorff:2002tj,Abe:2005wn}. 
The spectrum of the superparticles 
depends on whether they are in the bulk or on the boundary. 
If we do not assume any hierarchies among the 5D parameters, 
an $\cO(\mbox{TeV})$ KK scale is allowed 
only in the case that all the standard model fields 
are localized on the boundary. 
In the other cases, the gauginos become much heavier than the sfermions 
at the KK scale~$m_{\rm KK}$,  
and the spectrum becomes similar to that of the gaugino mediation  
when $m_{\rm KK}=\cO(10^{16}\mbox{GeV})$. 
We also provided some comments on cosmology.

\subsection*{Acknowledgements}
The authors would like to thank Hiroyuki Abe and 
Tetsutaro Higaki for useful information and discussions. 
This work was supported in part by 
Grant-in-Aid for Scientific Research (C) No.25400283  
from Japan Society for the Promotion of Science (Y.S.),   
and a Grant for Excellent Graduate Schools, MEXT, Japan (Y.Y.).

\appendix

\section{Equivalence to Scherk-Schwarz SUSY breaking} \label{SSbreaking}
Here we comment on an interpretation of the result in Sec.~\ref{Casimir_stb} 
from the viewpoint of the Scherk-Schwarz SUSY breaking, 
and compare (\ref{Veff:single0}) with the result 
in Ref.~\cite{vonGersdorff:2003rq}. 
For this purpose, we consider a case that no gaugino condensation term exists 
($A=0$) and there are $n_{H1}$ massless hypermultiplets and 
$n_{H2}$ hypermultiplets with a common bulk mass~$M_H$, 
and no antiperiodic fields (\ie, $n_a^A=n_V^A=0$). 
Then (\ref{Veff:single0}) becomes 
\bea
 V_{\rm eff} \eql -\frac{6\abs{\vph_C}^4\abs{W_0}^2}{\tau^4}
 \brc{\xi_1-\frac{\xi_2}{12}F\brkt{c_H\tau}}+\cdots \nonumber\\
 \eql -\frac{3\abs{\vph_C}^2\abs{F^T}^2}{2\tau^4}
 \brc{\xi_1-\frac{\xi_2}{12}F\brkt{c_H\tau}}+\cdots, 
 \label{Veff:single}
\eea
where the ellipsis denotes higher-order terms in $\xi_1$ or $\xi_2$ 
and $Q_a$-dependent terms, and 
\bea
 \xi_1 \defa \frac{(n_V-n_{H1}+2)\zeta(3)}{32\pi^2}, \;\;\;\;\;
 \xi_2 \equiv \frac{n_{H2}}{8\pi^2}, \nonumber\\
 F(x) \defa \zeta(3)\brc{3\cZ_P(x)-2x\cZ'_P(x)+\frac{x^2}{2}\cZ''_P(x)}. 
\eea
At the second equality, we have used that 
\bea
 F^T \eql \frac{2\bar{\vph}_C^2\bar{W_0}}{\vph_C}+\cdots,  \label{F^T:value}
\eea

It is well-known that SUSY breaking by $F^T$ is equivalent 
to the Scherk-Schwarz SUSY breaking as shown in 
Refs.~\cite{vonGersdorff:2002tj,Abe:2005wn,Marti:2001iw,Kaplan:2001cg}. 
In the latter mechanism, SUSY is broken 
by the twisted boundary condition, 
\be
 \Phi(x,y+2L) = e^{-2\pi i\vec{\omg}\cdot\vec{\sgm}}\Phi(x,y), 
 \label{SScond}
\ee
where $2L$ is the circumference of $S^1$, 
and $\Phi$ denotes a $SU(2)_R$-doublet field, \ie, 
the gravitino, the gaugino or the hyperscalar. 
From the consistency with the orbifold projection, 
the twist vector must be $\vec{\omg}=(\omg_1,\omg_2,0)$. 
We can always go to the periodic field basis by redefining fields as
$\Phi\to e^{i\vec{\omg}\cdot\vec{\sgm}f(y)}\Phi$, where a function $f(y)$ satisfies 
$f(y+2L) = f(y)+2\pi$. 
Then the radion F-term~$F^T$ is shifted by~\cite{Abe:2004ar,Abe:2005wn} 
\be
 F^T \to F^T+2\pi (\omg_2-i\omg_1)\abs{\vph_C}. 
\ee
Conversely, the nonzero value of $F^T$ in (\ref{F^T:value}) can be translated into 
the $SU(2)_R$ twist~\footnote{
This translation is possible only in the flat spacetime. 
In the warped spacetime, $SU(2)_R$-twisted boundary conditions 
lead to an inconsistency~\cite{Abe:2005wn,Hall:2003yc}. 
} 
with 
\be
 \vec{\omg} = \frac{1}{2\pi\abs{\vph_C}}
 \brkt{-\Im F^T,\Re F^T,0}, \label{VEV:F^T}
\ee
by the field redefinition~$\Phi\to e^{-i\vec{\omg}\cdot\vec{\sgm}f(y)}\Phi$.\footnote{
Especially, we can cancel 
the boundary constant superpotential~$W_0$ 
by choosing the function~$f(y)$ as a step-function~\cite{Abe:2005wn}. 
}

Using (\ref{VEV:F^T}) and (\ref{D-fixing}), the potential~(\ref{Veff:single}) 
is rewritten as 
\be
 V_{\rm eff} \simeq -\frac{6\pi^2\abs{\vec{\omg}}^2}{\tau^6}
 \brc{\xi_1-\frac{\xi_2}{12}F\brkt{c_H\tau}}+\cdots. \label{Veff:single3}
\ee
Now we assume that $c_H\tau\simgt 2$. 
Then, since 
\be
 \cZ_P(x) \simeq \frac{e^{-2x}}{\zeta(3)}\brkt{1+2x}, 
\ee
for $x\simgt 1$, the function~$F(x)$ is approximated as 
\be
 F(x) \simeq e^{-2x}\brkt{3+6x+6x^2+4x^3}.  \label{ap:F}
\ee
We find that (\ref{Veff:single3}) and (\ref{ap:F}) 
agrees with (3.6) in Ref.~\cite{vonGersdorff:2003rq} 
under the following identification. 
\bea
 \tau &\leftrightarrow & (\pi L)^{2/3}\phi^{1/3}, \nonumber\\
 c_H = M_H \Led^{1/3} &\leftrightarrow & M(L\pi)^{1/3}, 
 \nonumber\\
 (n_{H1},n_{H2},n_V) &\leftrightarrow & (N_h,N_H,N_V), \nonumber\\
 \abs{\vec{\omg}} &\leftrightarrow & \omg. 
\eea
The quantities in the right-hand side are the ones in Ref.~\cite{vonGersdorff:2003rq}. 
We have used that $M_5=\Led^{-1/3}$ in the unit of $M_{\rm Pl}$. 
In this comparison, we assumed that $\omg\ll 1$ and use the formula, 
\be
 \sum_{k=1}^\infty\frac{\sin^2(\pi\omg k)}{k^5} = \pi^2\zeta(3)\omg^2+\cO(\omg^4). 
\ee
From the relation~(\ref{ReT-Lphys}), we find that $V_{\rm eff}$ 
scales as $\Led^{-4}$, which is peculiar to the Casimir energy. 
In Ref.~\cite{vonGersdorff:2003rq}, the (negative) bulk cosmological constant 
and the brane tensions are introduced as counterterms to absorb 
the nonvanishing vacuum energy and ensure the 4D Minkowski spacetime. 
In the context of 5D SUGRA, 
the introduction of the 5D cosmological constant requires gauging some isometry 
by the graviphoton. 
As pointed out in Ref.~\cite{Abe:2005wn},
this gauging is inconsistent with the Scherk-Schwarz twisted boundary 
condition~(\ref{SScond}). 
Thus we do not introduce such counterterms here. 
Instead we assume the uplifting sector that consists of a chiral superfield~$X$ 
originating from a brane-localized chiral multiplet 
to achieve vanishing 4D cosmological constant. 

When 
\be
 \frac{1}{3} < \frac{\xi_2}{12\xi_1} 
 = \frac{n_{H2}}{3(n_V-n_{H1}+2)\zeta(3)} \simlt \cO(1), 
\ee
the potential~(\ref{Veff:single3}) has a minimum at $c_H\tau=\cO(1)$. 
Namely, the KK scale is 
\be
 m_{\rm KK} \equiv \frac{\pi}{\Led} = \cO(1)\times M_H, 
\ee 
and all the superparticles in the bulk have a common masses, 
\be
 m_{\rm SB} = \frac{2\pi\abs{\vec{\omg}}}{\Led}
 = \frac{2\abs{W_0}}{\Led},  
\ee
where we have used that $\abs{\vec{\omg}}=\abs{W_0}/\pi$. 
Thus, to obtain the TeV-scale superparticles, 
we have to assume a large hierarchy among the fundamental scales of 
the 5D theory~$M_5$, $M_H$ and $\abs{W_0}^{1/3}$ since 
\be
 \cO\brkt{\abs{W_0}M_H} = \cO\brkt{\abs{W_0}M_5^3} = \cO\brkt{10^{-15}}. 
 \label{rel:5Dparameters}
\ee

%%%%%%%%%%%%%%%%%%%%%%%%%%%% References %%%%%%%%%%%%%%%%%%%%%%%%%%%%%%

\end{document}